\documentclass[aps,pre,twocolumn,superscriptaddress]{revtex4-1}
\usepackage{graphicx}
\usepackage{dcolumn}
\usepackage{bm}
\usepackage{amsmath}
\usepackage{amssymb}
\usepackage{color}

\begin{document}

\title{Curvature sensing of curvature-inducing proteins with internal structure}

\author{Hiroshi Noguchi}
\email[]{noguchi@issp.u-tokyo.ac.jp}
\affiliation{Institute for Solid State Physics, University of Tokyo, Kashiwa, Chiba 277-8581, Japan}


\begin{abstract}
Many types of peripheral and transmembrane proteins can sense and generate membrane curvature.
Laterally isotropic proteins and crescent proteins with twofold rotational symmetry, 
such as Bin/Amphiphysin/Rvs superfamily proteins, have been studied theoretically.
However, proteins often have an asymmetric structure or a higher rotational symmetry.
We theoretically studied the curvature sensing of proteins with asymmetric structures and structural deformations.
First, we examined proteins consisting of two rod-like segments.
When proteins have mirror symmetry, their sensing ability is similar to that of single-rod proteins;
hence, with increasing protein density on a cylindrical membrane tube, second- or first-order transition occurs at a middle or small 
tube radius, respectively.
As asymmetry is introduced, this transition becomes a continuous change,
and metastable states appear at high protein densities.
Protein with threefold, fivefold, or higher rotational symmetry has laterally isotropic bending energy.
However, when a structural deformation is allowed, 
 the protein can have a preferred orientation and stronger curvature sensing.
\end{abstract}

\maketitle

\section{Introduction}

In living cells, biomembranes are primarily composed of lipids and proteins.
Transmembrane proteins span the membrane, while
peripheral proteins bind and unbind to the membrane surface.
Many of these proteins modify membrane properties, such as bending rigidity, spontaneous curvature, membrane thickness,
and viscosity.
Curvature-inducing proteins, such as Bin/Amphiphysin/Rvs (BAR) superfamily proteins,
regulate cell and organelle membrane shapes~\cite{mcma05,suet14}.
The BAR superfamily proteins have a crescent binding-domain (BAR domain), 
which is a dimer with twofold rotational symmetry.
The BAR domain bends membranes along its axis and
generates a cylindrical membrane tube~\cite{mcma05,suet14,joha15,itoh06,masu10,mim12a,fros08}.
Clathrin and coat protein molecules assemble to form  spherical cargo, generating spherical membrane buds~\cite{joha15,hurl10,mcma11,bran13,mett18,tayl22}.
These curvature-inducing proteins sense membrane curvature
and are concentrated at the membrane locations of their preferred curvatures.
Curvature sensing of
BAR proteins~\cite{baum11,has21,sorr12,prev15,tsai21}, dynamin~\cite{roux10},  annexins~\cite{more19},
G-protein coupled receptors (GPCRs)~\cite{rosh17}, ion channels~\cite{aimo14,yang22}, and Ras proteins~\cite{lars20} 
 has been reported using tethered vesicles.
The dependence of protein binding on vesicle size also indicates curvature sensing~\cite{lars20,hatz09,zeno19}.

\begin{figure}[tbh]
\includegraphics[width=8.5cm]{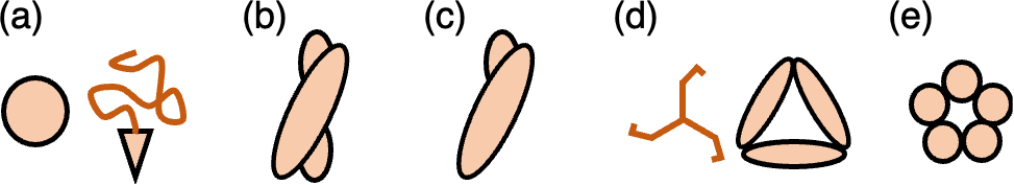}
\caption{
 Lateral symmetries of proteins on a membrane.
(a) Laterally isotropic proteins, modeled as a circular shape on the membrane.
Intrinsically disordered protein (IDP) domains and the insertion of a hydrophobic $\alpha$-helix
can bend the membrane isotropically.
(b) Twofold rotational symmetry. BAR superfamily proteins form a dimer that has twofold symmetry.
(c) Asymmetric proteins. Dynamin and amphipathic peptides such as melittin do not have a rotational symmetry.
(d) Threefold rotational symmetry.  The clathrin monomer has a threefold symmetric shape (left panel).
The trimers of proteins such as annexin and microbial rhodopsins also have threefold symmetry (right panel).
(e) Fivefold rotational symmetry. Transmembrane proteins, such as ion channels, and their assemblies 
often have fivefold or higher symmetries.
}
\label{fig:crot}
\end{figure}

Theoretically, curvature-inducing proteins have been modeled as laterally isotropic or crescent objects.
For isotropic objects, the Canham--Helfrich model~\cite{canh70,helf73} was applied to the bending energy~\cite{prev15,tsai21,nogu22a,gout21,nogu21a,nogu21b}. 
For crescent objects, anisotropic bending energies were considered~\cite{nogu22a,four96,kral99,akab11,tozz21,nogu22,nogu23b}.
An elliptical shape was typically considered, such that a twofold rotational and mirror-symmetric shape was assumed.
However, actual proteins often have more complicated shapes.
BAR domains have twofold rotational symmetry but are chiral and are not mirror symmetric (see Fig.~\ref{fig:crot}(b)).
Their chirality is the origin of the helical assembly of the BAR domains~\cite{mim12a,fros08} 
and is important for generating membrane tubes with a constant radius~\cite{nogu19a}.
Many of BAR and other curvature-inducing proteins have intrinsically disordered domains~\cite{piet13},
and recent experiments have demonstrated that these disordered domains have significant effects on curvature generation~\cite{busc15,zeno19,snea19}.
Theoretically, they are treated as excluded-volume linear polymer chains.
At a low polymer density on the membrane surface, 
polymer--membrane interactions can weakly induce a spontaneous curvature in a laterally isotropic manner~\cite{hier96,bick01,auth03,auth05,wu13}.
Conversely, at high densities, inter-polymer interactions can induce a large spontaneous curvature~\cite{hier96,wu13,mars03,evan03a,wern10}
and promote  membrane tubulation or prevent it because of the repulsion between polymers~\cite{nogu22b}. 

\begin{figure}[tbh]
\includegraphics[]{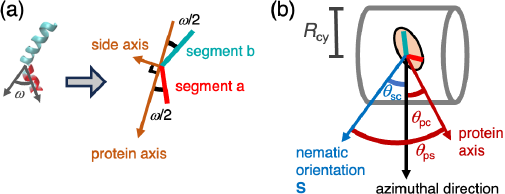}
\caption{
Schematic of an asymmetric curvature-inducing protein.
(a) Model of the protein with two rod-like segments.
(b) Protein on a cylindrical membrane. The angles between the nematic direction {\bf S}, azimuthal direction, and/or protein axis are depicted.
}
\label{fig:cart}
\end{figure}

In this study, we consider two types of curvature-inducing proteins: 
asymmetric proteins, and proteins with threefold or higher rotational symmetry (see Fig.~\ref{fig:crot}).
Dynamin~\cite{ferg12,anto16,pann18} has an asymmetric structure,
and its helical assembly induces membrane fission by choking a membrane neck.
Melittin and amphipathic peptides~\cite{sato06,rady17,guha19,miya22}
bind onto the membrane, and their circular assembly forms a membrane pore.
G{\'o}mez-Llobregat et al. reported the curvature sensing of three amphipathic peptides
using a coarse-grained simulation of a buckled membrane~\cite{gome16}.
They revealed that melittin and the amphipathic peptides LL-37 (PDB: 2k6O) exhibited asymmetric curvature sensing,
which means the angle distribution with respect to the buckled axis was not symmetric.
We use a protein model consisting of two crescent-rod-like segments connected by a kink, like melittin (see Fig.~\ref{fig:cart}(a)),
and investigate how the asymmetry modifies curvature sensing.

Many transmembrane proteins, such as ion channels~\cite{tray10,syrj21} and GPCRs~\cite{erns14,naga21,venk13,shib18},
form rotational symmetric structures.
Several types of microbial rhodopsins form a trimer or pentamer with three- or fivefold symmetry, respectively~\cite{shib18}.
Moreover, peripheral proteins can have threefold symmetry.
For example, the clathrin monomer has threefold symmetry~\cite{hurl10}, and 
annexin A5 molecules form a trimer with a triangular shape~\cite{gerk05,olin00}.
Recently, deformation of the lipid bilayer induced by the hydrophobic mismatch of rotationally symmetric transmembrane proteins
was theoretically studied~\cite{alas23}.
In this study, we investigate curvature sensing of $N$-fold rotationally symmetric proteins with $N\ge 3$.
The rigid rotationally symmetric proteins exhibit isotropic bending energy.
However, the anisotropy can be induced by protein deformation.

The previous theoretical models of curvature-inducing proteins are outlined in Sec.~\ref{sec:1rod}.
The curvature sensing of asymmetric proteins is described in Sec.~\ref{sec:2rod}.
The protein model is presented in Sec.~\ref{sec:2rodmodel}.
Curvature sensing at low-density limits and at finite densities is described in  Sec.~\ref{sec:2rodsin} and \ref{sec:2rodden}, respectively.
Sec.~\ref{sec:three} discusses proteins with threefold or higher rotational symmetries.
Sec.~\ref{sec:sum} concludes the paper.

\section{Protein models with anisotropic bending energy}\label{sec:1rod}

Crescent proteins were modeled to have different bending rigidities and spontaneous curvatures along
the protein axis and in the perpendicular (side) direction.
Note that this protein axis is set along the main preferred curvature of the protein on the membrane,
so that it can be different from the protein axis of the elliptical approximation (e.g., BAR-PH domains~\cite{masu10,mim12a}).
The membrane curvatures along these two directions are given by
\begin{eqnarray}
C_{\ell 1} &=& C_1\cos^2(\theta_{\mathrm {pc}}) +C_2\sin^2(\theta_{\mathrm {pc}}) = H + D\cos(2\theta_{\mathrm {pc}}), \ \ \ \\
C_{\ell 2} &=& C_1\sin^2(\theta_{\mathrm {pc}}) +C_2\cos^2(\theta_{\mathrm {pc}}) = H - D\cos(2\theta_{\mathrm {pc}}), 
\end{eqnarray}
where $\theta_{\mathrm {pc}}$ is the angle between the protein axis and direction of either principal membrane curvature
(the azimuthal direction is chosen for a cylindrical membrane as depicted in Fig.~\ref{fig:cart}(b)).
$H=(C_1+C_2)/2$ and $D= (C_1-C_2)/2$ represent the mean and deviatoric curvatures of the membrane, respectively,
where $C_1$  and $C_2$ represent the principal curvatures.
The bending energy of a protein is expressed as \cite{nogu22a,nogu16,nogu22}
\begin{eqnarray} \label{eq:1rod1}
U_{\mathrm {1rod}} &=& \frac{\kappa_{\mathrm {p}}a_{\mathrm {p}}}{2}(C_{\ell 1} - C_{\mathrm {p}})^2 + \frac{\kappa_{\mathrm {s}}a_{\mathrm {p}}}{2}(C_{\ell 2} - C_{\mathrm {s}})^2 \\ \nonumber
&=&\ a_{\mathrm {p}}\bigg\{ \frac{(\kappa_{\mathrm {p}}+ \kappa_{\mathrm {s}})}{2}\bigg[ H^2  +  \frac{D^2}{2}(\cos(4\theta_{\mathrm {pc}})+1) \bigg] \\ \nonumber
&& -  (\kappa_{\mathrm {p}}C_{\mathrm {p}}+\kappa_{\mathrm {s}}C_{\mathrm {s}})H + \frac{\kappa_{\mathrm {p}}C_{\mathrm {p}}^2+\kappa_{\mathrm {s}}C_{\mathrm {s}}^2}{2}  \\ \nonumber
&& +  (\kappa_{\mathrm {p}}-\kappa_{\mathrm {s}})HD\cos(2\theta_{\mathrm {pc}}) \\ 
&& - (\kappa_{\mathrm {p}}C_{\mathrm {p}}-\kappa_{\mathrm {s}}C_{\mathrm {s}})D\cos(2\theta_{\mathrm {pc}}) \bigg\}, \label{eq:1rod2}
\end{eqnarray}
where $a_{\mathrm {p}}$ is the contact area of the bound protein, $\kappa_{\mathrm {p}}$ and $C_{\mathrm {p}}$ are the bending rigidity and spontaneous curvature along the protein axis, respectively,
and $\kappa_{\mathrm {s}}$ and $C_{\mathrm {s}}$ are along the side axis.
From the comparison of the experimental data of tethered vesicles~\cite{prev15,tsai21}, 
the bending rigidity and spontaneous curvature along the protein axis were estimated:
$\kappa_{\mathrm{p}}/k_{\mathrm{B}}T= 82\pm 20$ and $C_{\mathrm{p}}(\mathrm{nm}^{-1}) = -0.047 + 0.0003(\kappa_{\mathrm{p}}/k_{\mathrm{B}}T-82) \pm 0.001$
for I-BAR domain, and
$30\lesssim\kappa_{\mathrm{p}}/k_{\mathrm{B}}T \lesssim 60$ 
and $0.06 \lesssim C_{\mathrm{p}}(\mathrm{nm}^{-1})\lesssim 0.09$ for N-BAR domain~\cite{nogu23b}.

Different forms of the anisotropic bending energy have also been used.
In Ref.~\citenum{four96}, only the linear terms of $H$ and $D$ were considered
in addition to the tilt energy.
In Ref.~\citenum{kral99}, the energy was considered to be
\begin{eqnarray} \label{eq:igl}
U_{\mathrm {grad}} &=& \frac{k_{\mathrm {m}}}{2}(H - H_0)^2 \\ \nonumber 
&& + \frac{k_{\mathrm {m}}+k_{\mathrm {d}}}{4}\big(D^2 - 2DD_0\cos(2\theta_{\mathrm {pc}})  + D_{0}^2\big).
\end{eqnarray}
The second term assumes an energy proportional to a rotational average in the squared gradient of the normal curvature, $C_\ell-C_p$, with respect to the protein rotation.
In this form, the protein depends only weakly on the protein orientation; the cross term of $HD$ does not appear and the $D^2$ term is independent of the angle $\theta_{\mathrm {pc}}$.

In these protein models, the bending energy depends on the angle only as a function of $\cos(2\theta_{\mathrm {pc}})$, owing to symmetry.
For asymmetric proteins, the energy can include an odd function of the angle $\theta_{\mathrm {pc}}$.
To the best of our knowledge, such a term was previously considered only in the model by Akabori and Santangelo~\cite{akab11}.
They added the following term to Eq.~(\ref{eq:1rod1}):
\begin{equation} \label{eq:aka}
U_{\mathrm {asy}} = k_{\mathrm {asy}}( D\sin(2\theta_{\mathrm {pc}}) - C_{\mathrm {asy}})^2,
\end{equation}
where $D\sin(2\theta_{\mathrm {pc}})$ is  the non-diagonal element of the curvature tensor. 
In Ref.~\citenum{gome16}, this model was used to estimate the bending rigidities of amphipathic peptides.
However, this model does not have a microscopic basis.
In this study, we examine the bending energies of asymmetric proteins using a 2-rod protein model.

\section{Protein consisting of two rods}\label{sec:2rod}

\subsection{Protein Model}\label{sec:2rodmodel}

We consider a protein or peptide consisting of two segments (segments $a$ and $b$ in Fig.~\ref{fig:cart}(a)).
Each segment is modeled as the symmetric protein model (in the absence of side bending rigidity for simplicity),
and the orientations of the two segments have an angle $\omega$ on the membrane surface.
Melittin is an example of this type of molecule, 
in which two alpha helices are connected by a kink.
The bending energy of one protein is expressed as
\begin{eqnarray}
U_{\mathrm {2rod}} &=&\   \frac{\kappa_{\mathrm {pa}}a_{\mathrm {pa}}}{2}(C_{\ell 1{\mathrm {a}}} - C_{\mathrm {pa}})^2 + \frac{\kappa_{\mathrm {pb}}a_{\mathrm {pb}}}{2}(C_{\ell 1{\mathrm {b}}} - C_{\mathrm {pb}})^2 \nonumber \\ \label{eq:u2}
&=&\   \kappa_{\mathrm {pm}}a_{\mathrm {p}}\bigg[ (H - C_{\mathrm {pm}})^2 + C_{\mathrm {pd}}^2  \\ \nonumber
&&+ 2(H - C_{\mathrm {pm}})D\cos(\omega)\cos(2\theta_{\mathrm {pc}})  \\ \nonumber
&&+ 2C_{\mathrm {pd}}D\sin(\omega)\sin(2\theta_{\mathrm {pc}}) \\ \nonumber
&&+ \frac{D^2}{2}(\cos(2\omega)\cos(4\theta_{\mathrm {pc}})+1)     \bigg]  \\ \nonumber
&&+ \kappa_{\mathrm {pd}}a_{\mathrm {p}}\bigg[ -2HC_{\mathrm {pd}} +2C_{\mathrm {pm}}C_{\mathrm {pd}} \\ \nonumber
&&-2C_{\mathrm {pd}}D\cos(\omega)\cos(2\theta_{\mathrm {pc}}) \\ \nonumber
&&-2(H-C_{\mathrm {pm}})D\sin(\omega)\sin(2\theta_{\mathrm {pc}}) \\ \nonumber
&&- \frac{D^2}{2}\sin(2\omega)\sin(4\theta_{\mathrm {pc}}) \bigg],
\end{eqnarray}
where $C_{\mathrm {pm}}=(C_{\mathrm {pa}}+C_{\mathrm {pb}})/2$,  $C_{\mathrm {pd}}=(C_{\mathrm {pa}}-C_{\mathrm {pb}})/2$,
$\kappa_{\mathrm {pm}}a_{\mathrm {p}}=(\kappa_{\mathrm {pa}}a_{\mathrm {pa}}+\kappa_{\mathrm {pb}}a_{\mathrm {pb}})/2$, and
$\kappa_{\mathrm {pd}}a_{\mathrm {p}}=(\kappa_{\mathrm {pa}}a_{\mathrm {pa}}-\kappa_{\mathrm {pb}}a_{\mathrm {pb}})/2$.
We use $\kappa_{\mathrm {pm}}=50k_{\mathrm {B}}T$ and $a_{\mathrm {p}}C_{\mathrm {pm}}^2=0.1$.
These values are typical of curvature-inducing proteins.
The angle $\omega=\pi/6$ is used, unless otherwise specified.
Note that $\kappa_{\mathrm {pd}}$ varies according to the bending rigidity difference and the area difference between the two segments.

In Eq.~(\ref{eq:u2}), the deviatoric curvature $D$ and angle $\theta_{\mathrm {pc}}$ always appear as pairs
as a function of $D\cos(2\theta_{\mathrm {pc}})$ and/or $D\sin(2\theta_{\mathrm {pc}})$.
The asymmetric terms $\propto HD\sin(2\theta_{\mathrm {pc}})$ and $\propto D^2\sin(4\theta_{\mathrm {pc}})$
exist in addition to the term $\propto D\sin(2\theta_{\mathrm {pc}})$.
Therefore, the asymmetric energy described in Eq.~(\ref{eq:aka})~\cite{akab11} is insufficient to express the asymmetric bending energy.

For a symmetric protein ($C_{\mathrm {pd}}=k_{\mathrm {pd}}=0$), the bending energy is expressed as
\begin{eqnarray}
U_{\mathrm {2rod}}^{\mathrm {sym}} &=& \frac{\kappa_{\mathrm {pm}}a_{\mathrm {p}}}{2}(1+\cos(\omega))(C_{\ell 1} - C_{\mathrm {pm}})^2 \nonumber \\
&&\ + \frac{\kappa_{\mathrm {pm}}a_{\mathrm {p}}}{2}(1-\cos(\omega))(C_{\ell 2} - C_{\mathrm {pm}})^2 \nonumber \\
&&\ - \frac{\kappa_{\mathrm {pm}}a_{\mathrm {p}}}{2}D^2(1-\cos(2\omega))\cos(4\theta_{\mathrm {pc}}). \label{eq:u2sym}
\end{eqnarray}
The first and second terms correspond to the bending energies along the main and side axes of the protein in Eq.~(\ref{eq:1rod1}), respectively.
However, the last term is new.
At $\omega =0$, the second and last terms vanish,
and with increasing $\omega$, they increase.

\begin{figure}[tbh]
\includegraphics[]{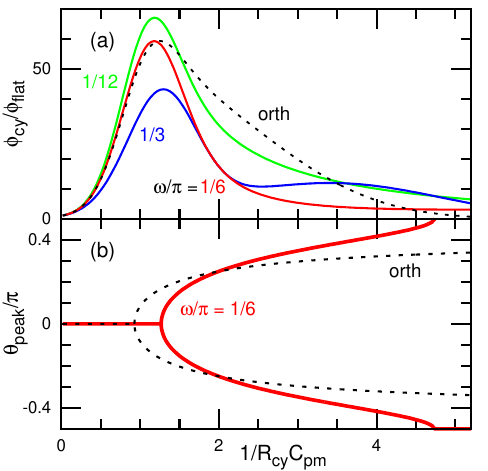}
\caption{
Binding of symmetric proteins ($\kappa_{\mathrm{pd}}=C_{\mathrm{pd}}=0$) at the low-density limit.
(a) Binding density $\phi_{\mathrm{cy}}$ on a cylindrical membrane  with respect to 
the density $\phi_{\mathrm{flat}}$ on a flat membrane.
The solid lines represent the data for $\omega/\pi= 1/12$, $1/6$, and $1/3$
(from top to bottom in the left region, respectively).
(b) Peak position of the angle $\theta_{\mathrm{pc}}$ at $\omega/\pi= 1/6$.
The dashed lines in (a) and (b) represent the data obtained using the orthogonal approximation at  $\omega/\pi= 1/6$.
}
\label{fig:scd0bd0}
\end{figure}

\subsection{Isolated Proteins}\label{sec:2rodsin}

\begin{figure}[tbh]
\includegraphics[]{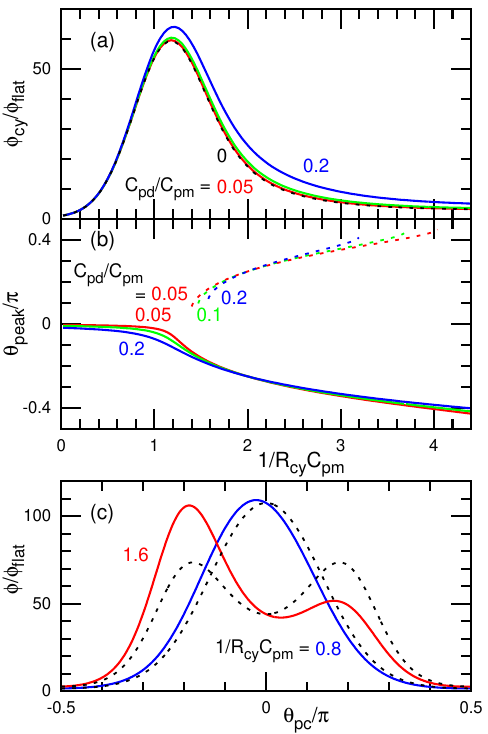}
\caption{
Binding of asymmetric proteins with $\kappa_{\mathrm{pd}}=0$ and $\omega/\pi= 1/6$ at the low-density limit.
(a) Binding density $\phi_{\mathrm{cy}}$ on a cylindrical membrane with respect to 
the density $\phi_{\mathrm{flat}}$ on a flat membrane.
The solid lines represent the data at $C_{\mathrm{pd}}/C_{\mathrm{pm}}=0.2$, $0.1$, and $0.05$
from top to bottom.
The dashed line represents the data at $C_{\mathrm{pd}}=0$ (the symmetric condition).
(b) Peak position of the angle $\theta_{\mathrm{pc}}$.
The solid and dashed lines represent the first and second peaks, respectively.
(c) Distribution of the angle $\theta_{\mathrm{pc}}$ at $1/R_{\mathrm{cy}}C_{\mathrm{pm}}=0.8$ and $1.6$.
The solid and dashed lines represent the data for $C_{\mathrm{pd}}/C_{\mathrm{pm}}=0.1$ and $0$, respectively.
}
\label{fig:scds}
\end{figure}

First, we consider protein binding at the low-density limit,
in which bound proteins are isolated on a membrane and inter-protein interactions are negligible.
Hence, the density $\phi$ of bound proteins is given by $\phi= (1/2\pi)\int_{-\pi}^{\pi} \exp[\beta(\mu-U_{\mathrm {2rod}})]\ {\mathrm d}\theta_{\mathrm {pc}}$,
where $\mu$ is the binding chemical potential and $\beta=1/k_{\mathrm {B}}T$.
The binding ratio of proteins to a cylindrical membrane tube with respect to a flat membrane
is expressed as 
\begin{equation}\label{eq:phi0}
\frac{\phi_{\mathrm {cy}}}{\phi_{\mathrm {flat}}} = \frac{\exp(\beta U_{\mathrm {2rod}}^{\mathrm {flat}})}{2\pi}\int_{-\pi}^{\pi} \exp(-\beta U_{\mathrm {2rod}}^{\mathrm {cy}})\ {\mathrm d}\theta_{\mathrm {pc}},
\end{equation}
where $U_{\mathrm {2rod}}^{\mathrm {flat}}$ is the bending energy for the flat membrane ($H=D=0$) and 
$U_{\mathrm {2rod}}^{\mathrm {cy}}$ is that for the cylindrical membrane ($H=D=1/2R_{\mathrm {cy}}$).
This ratio $\phi_{\mathrm {cy}}/\phi_{\mathrm {flat}}$ is independent of $\mu$ at the low-density limit ($\phi_{\mathrm {cy}}\ll 1$ and $\phi_{\mathrm {flat}} \ll 1$).

\begin{figure}[tbh]
\includegraphics[]{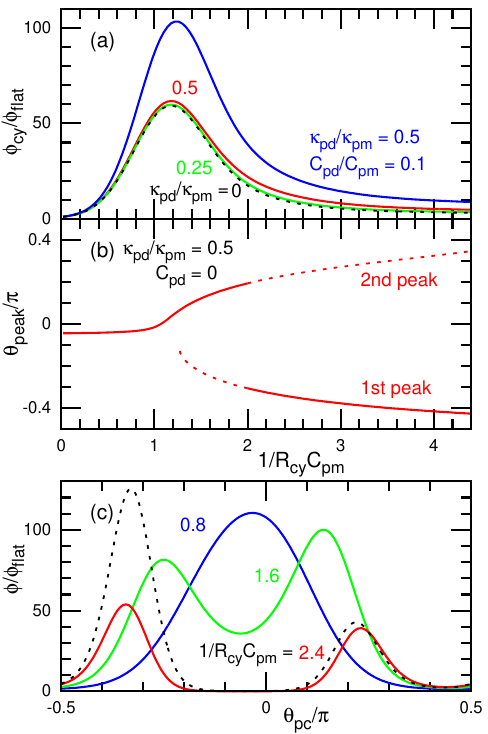}
\caption{
Binding of asymmetric proteins with $\kappa_{\mathrm{pd}}> 0$ and $\omega/\pi= 1/6$ at the low-density limit.
(a) Binding density $\phi_{\mathrm{cy}}$ on a cylindrical membrane with respect to 
the density $\phi_{\mathrm{flat}}$ on a flat membrane.
The uppermost line represents the data at  $\kappa_{\mathrm{pd}}/\kappa_{\mathrm{pm}}=0.5$ and $C_{\mathrm{pd}}/C_{\mathrm{pm}}=0.1$.
The lower two solid lines represent the data for $\kappa_{\mathrm{pd}}/\kappa_{\mathrm{pm}}=0.5$ and $0.25$ at $C_{\mathrm{pd}}=0$.
The dashed line represents the data at $\kappa_{\mathrm{pd}}=0$ and $C_{\mathrm{pd}}=0$ (the symmetric condition).
(b) Peak position of the angle $\theta_{\mathrm{pc}}$ at $\kappa_{\mathrm{pd}}/\kappa_{\mathrm{pm}}=0.5$ and $C_{\mathrm{pd}}=0$.
The solid and dashed lines represent the first and second peaks, respectively.
(c) Distribution of the angle $\theta_{\mathrm{pc}}$ at $\kappa_{\mathrm{pd}}/\kappa_{\mathrm{pm}}=0.5$.
The solid lines represent the data for $1/R_{\mathrm{cy}}C_{\mathrm{pm}}=0.8$, $1.6$, and $2.4$ at $C_{\mathrm{pd}}=0$.
The dashed line represents the data at $1/R_{\mathrm{cy}}C_{\mathrm{pm}}=2.4$ and $C_{\mathrm{pd}}/C_{\mathrm{pm}}=0.1$.
}
\label{fig:sbds}
\end{figure}

Figure~\ref{fig:scd0bd0} shows the dependence on the curvature $1/R_{\mathrm {cy}}$ of the cylindrical membrane
for symmetrical proteins (Eq.~(\ref{eq:u2sym})) with a fixed angle $\omega$.
The binding density reaches a maximum at $1/R_{\mathrm {cy}}C_{\mathrm {pm}}\simeq 1.2$, 
and the maximum level decreases with increasing $\omega$.
Hence, the preferred curvature of the protein is slight higher than that of each segment, $C_{\mathrm {pm}}$.
The density distribution is mirror symmetric with respect to $\theta_{\mathrm {pc}}=0$ and has one or two peaks ($\theta_{\mathrm {peak}}$) at low or high membrane curvatures, respectively (see Fig.~\ref{fig:scd0bd0}(b) and the dashed lines in Fig.~\ref{fig:scds}(c)). 
This peak split occurs since the membrane curvature becomes higher than the preferred curvature for the protein
at high curvatures.
Each protein segment has the lowest bending energy when it is along the azimuthal direction for $1/R_{\mathrm {cy}}C_{\mathrm {pm}} \leq 1$, whereas it deviates from the azimuthal direction as $\theta_{\mathrm {pc}} \pm \omega/2 = \pm \arccos(\sqrt{R_{\rm cy}C_{\mathrm {pm}}})$ for $1/R_{\mathrm {cy}}C_{\mathrm {pm}} > 1$.
For $\omega=\pi/6$, the split point is shifted to a slightly higher membrane curvature (see Fig.~\ref{fig:scd0bd0}(b)),
since  two segments are tilted with $\pm\omega/2$,
when the protein is oriented in the azimuthal direction ($\theta_{\mathrm {pc}}=0$).
When the orthogonal protein model given in Eq.~(\ref{eq:1rod1}) is used (i.e., the last term in Eq.~(\ref{eq:u2sym}) is not accounted for),
the protein behavior can be reproduced well at low  membrane curvatures but not at high curvatures (see the dashed lines in Fig.~\ref{fig:scd0bd0}). 
Therefore, the last term in Eq.~(\ref{eq:u2sym}) significantly modifies protein behavior at high membrane curvatures.

Next, we consider the asymmetric proteins with $\omega=\pi/6$ (see Figs.~\ref{fig:scds} and \ref{fig:sbds}).
Figure~\ref{fig:scds} shows the case in which the spontaneous curvatures of two segments are different while keeping $\kappa_{\mathrm{pd}}=0$.
Since segment $a$ has a large spontaneous curvature, it is more oriented in the azimuthal direction than segment $b$.
Hence, the peak angle of $\theta_{\mathrm {pc}}$ becomes negative and decreases continuously with increasing $1/R_{\mathrm {cy}}$ (see Fig.~\ref{fig:scds}(b)). The upper peak becomes the second maximum for a finite range of $1/R_{\mathrm {cy}}$ (see the solid lines in Fig.~\ref{fig:scds}(c)).
The width of this range decreases with increasing $C_{\mathrm{pd}}$ (see dashed lines in Fig.~\ref{fig:scds}(b)).
However, the binding protein ratio $\phi_{\mathrm {cy}}/\phi_{\mathrm {flat}}$ is only slightly modified (see Fig.~\ref{fig:scds}(a)).

When the bending rigidities of the two segments are different, the proteins exhibit more complicated behavior.
For a small curvature of $1/R_{\mathrm {cy}}$, the angle distribution is slightly asymmetric and has a peak at $\theta_{\mathrm {pc}}<0$,
as in the previous case (compare Figs.~\ref{fig:scds}(c) and \ref{fig:sbds}(c)).
However, the peak position shifts to $\theta_{\mathrm {pc}}>0$ with increasing $1/R_{\mathrm {cy}}$,
and a second peak appears at $\theta_{\mathrm {pc}}<0$.
At $1/R_{\mathrm {cy}}C_{\mathrm {pm}} > 2$, the peak at $\theta_{\mathrm {pc}}<0$ becomes larger than the other one (see Fig.~\ref{fig:sbds}(b) and (c)).
These peak behaviors are caused by the last two terms in Eq.~(\ref{eq:u2}).
The sign of the penultimate term changes at $1/R_{\mathrm {cy}}C_{\mathrm {pm}} = 2$,
and the increase in $\theta_{\mathrm {peak}}$ at $1/R_{\mathrm {cy}}C_{\mathrm {pm}} \simeq 1$ is mainly due to the last term.

\begin{figure}[tbh]
\includegraphics[]{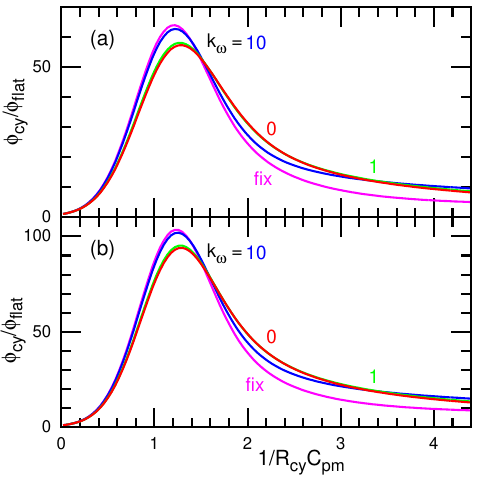}
\caption{
Binding density of asymmetric proteins with the harmonic angle potential at the low-density limit. 
The potential strength is varied as $k_{\omega}= 0$, $1$, and $10$ at $\omega_0/\pi= 1/6$.
The lowest lines in the right region ($1/R_{\mathrm{cy}}C_{\mathrm{pm}}>2$) 
represent the data when the angle is fixed at $\omega/\pi= 1/6$.
(a)  $\kappa_{\mathrm{pd}}=0$ and $C_{\mathrm{pd}}/C_{\mathrm{pm}}=0.2$.
(b)  $\kappa_{\mathrm{pd}}/\kappa_{\mathrm{pm}}=0.5$ and $C_{\mathrm{pd}}/C_{\mathrm{pm}}=0.1$.
}
\label{fig:har}
\end{figure}

When both the bending rigidities and spontaneous curvatures of the two segments are different,
the ratio $\phi_{\mathrm {cy}}/\phi_{\mathrm {flat}}$ can vary considerably from that of symmetric protein,
and the angle distribution can be more asymmetrical (see the uppermost line in Fig.~\ref{fig:sbds}(a) and the dashed line in Fig.~\ref{fig:sbds}(c)).
This increases in $\phi_{\mathrm {cy}}/\phi_{\mathrm {flat}}$ is due to the enhancement of protein curvature induction by the  effectively large
protein curvature  
($\kappa_{\mathrm {pa}}a_{\mathrm {pa}}C_{\mathrm {pa}} + \kappa_{\mathrm {pb}}a_{\mathrm {pb}}C_{\mathrm {pb}}= (\kappa_{\mathrm {pm}}a_{\mathrm {p}}C_{\mathrm {pm}} + \kappa_{\mathrm {pd}}a_{\mathrm {p}}C_{\mathrm {pd}})/2$).

Further, we consider the conformational fluctuations in the protein.
To allow an angle fluctuation of $\omega$,
a harmonic potential $U_{\omega}= (k_{\omega}k_{\mathrm {B}}T/2)(\omega-\omega_0)^2$ is added, where $\omega_0=\pi/6$.
At $k_{\omega}=0$, the two segments act as two separate rods, and 
the binding ratio $\phi_{\mathrm {cy}}/\phi_{\mathrm {flat}}$ exhibits a smaller peak and broader tail,
since the effective bending rigidity is smaller but the orientation is less constrained, respectively (see Fig.~\ref{fig:har}).
As $k_{\omega}$ increases, the ratio continuously changes into that at the fixed angle.

\begin{figure}[tbh]
\includegraphics[]{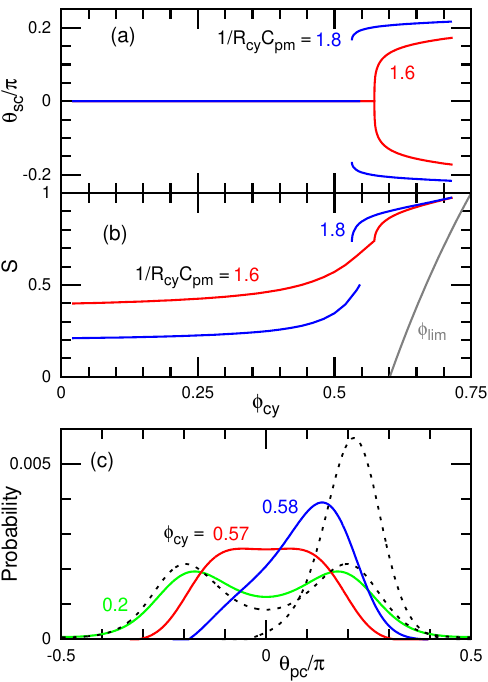}
\caption{
Binding of symmetric proteins ($\kappa_{\mathrm{pd}}=C_{\mathrm{pd}}=0$) for finite densities $\phi_{\mathrm{cy}}$ at $\omega/\pi= 1/6$.
The second- and first-order transitions occur at $1/R_{\mathrm {cy}}C_{\mathrm {pm}}=1.6$ and $1.8$, respectively.
(a) Angle $\theta_{\mathrm{sc}}$ between the orientational order and azimuthal direction.
(b) Orientational degree $S$ of the proteins.
The right line represents the maximum density $\phi_{\rm lim}(S)$.
(c) Distribution of the angle $\theta_{\mathrm{pc}}$.
The solid lines represent the data for $\phi_{\mathrm{cy}}=0.2$, $0.57$, and $0.58$ at $1/R_{\mathrm {cy}}C_{\mathrm {pm}}=1.6$.
The dashed lines represent the data for $\phi_{\mathrm{cy}}=0.5$ and $0.6$ at $1/R_{\mathrm {cy}}C_{\mathrm {pm}}=1.8$.
}
\label{fig:dcd0bd0}
\end{figure}

\begin{figure}[tbh]
\includegraphics[]{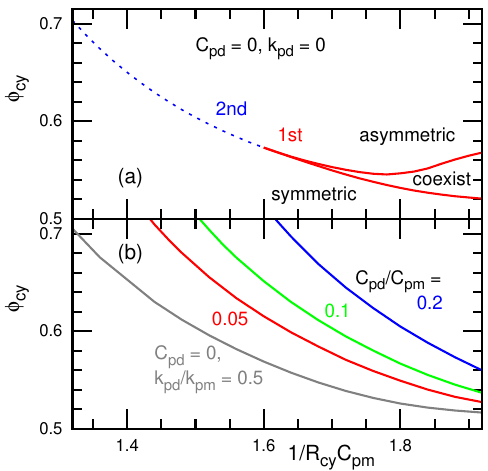}
\caption{
Phase diagram for (a) symmetric proteins and (b) asymmetric proteins.
(a) The dashed line represents the phase boundary of the second-order transition.
The proteins exhibit symmetric and asymmetric distributions with respect to $\theta_{\mathrm{sc}}=0$ below and above the lines, respectively.
Two states coexist between two solid lines.
(b) Boundaries of the metastable states.
The upper three lines represent the data for $C_{\mathrm{pd}}/C_{\mathrm{pm}}=0.2$, $0.1$, and $0.05$ at $\kappa_{\mathrm{pd}}=0$, from top to bottom.
The lowest line represents the data for $\kappa_{\mathrm{pd}}/\kappa_{\mathrm{pm}}=0.5$ and $C_{\mathrm{pd}}=0$.
}
\label{fig:bd}
\end{figure}

\begin{figure}[tbh]
\includegraphics[]{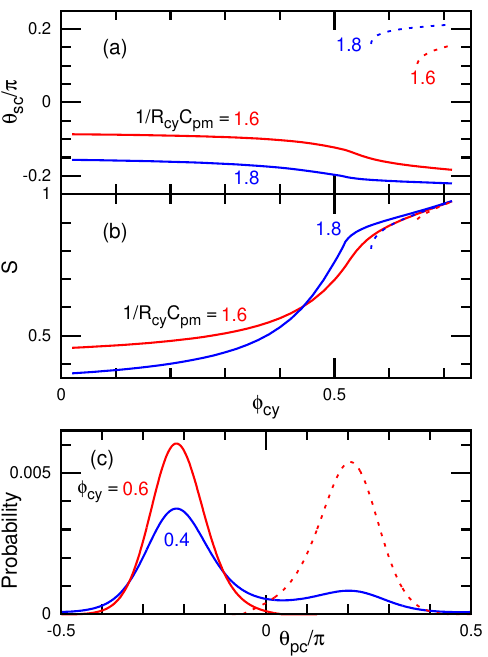}
\caption{
Binding of asymmetric proteins with $C_{\mathrm{pd}}/C_{\mathrm{pm}}=0.1$, $\kappa_{\mathrm{pd}}=0$, and $\omega/\pi= 1/6$ at finite densities $\phi_{\mathrm{cy}}$.
(a) Angle $\theta_{\mathrm{sc}}$ between the orientational order and azimuthal direction at $1/R_{\mathrm {cy}}C_{\mathrm {pm}}=1.6$ and $1.8$.
(b) Orientational degree $S$ of the proteins at $1/R_{\mathrm {cy}}C_{\mathrm {pm}}=1.6$ and $1.8$.
(c) Distribution of the angle $\theta_{\mathrm{pc}}$ for $\phi_{\mathrm{cy}}=0.4$ and $0.6$ at $1/R_{\mathrm {cy}}C_{\mathrm {pm}}=1.8$.
The solid and dashed lines represent the equilibrium and metastable states, respectively.
}
\label{fig:dcd1bd0}
\end{figure}

\begin{figure}[tbh]
\includegraphics[]{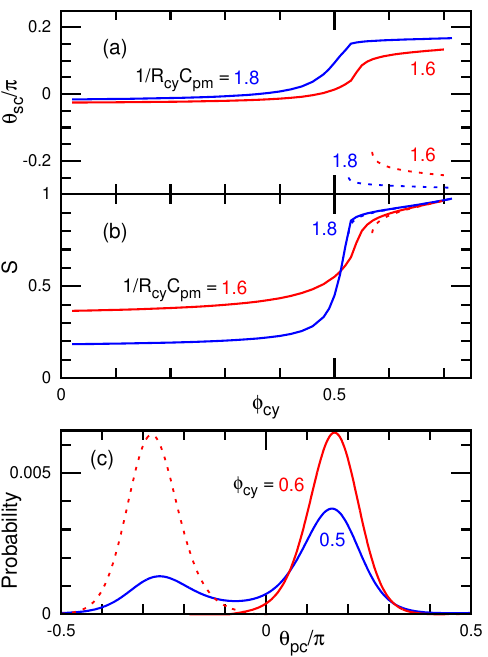}
\caption{
Binding of asymmetric proteins with $\kappa_{\mathrm{pd}}/\kappa_{\mathrm{pm}}=0.5$, $C_{\mathrm{pd}}=0$, and $\omega/\pi= 1/6$ at finite densities $\phi_{\mathrm{cy}}$.
(a) Angle $\theta_{\mathrm{sc}}$ between the orientational order and azimuthal direction at $1/R_{\mathrm {cy}}C_{\mathrm {pm}}=1.6$ and $1.8$.
(b) Orientational degree $S$ of the proteins at $1/R_{\mathrm {cy}}C_{\mathrm {pm}}=1.6$ and $1.8$.
(c) Distribution of the angle $\theta_{\mathrm{pc}}$ for $\phi_{\mathrm{cy}}=0.5$ and $0.6$ at $1/R_{\mathrm {cy}}C_{\mathrm {pm}}=1.8$.
The solid and dashed lines represent the equilibrium and metastable states, respectively.
}
\label{fig:dcd0bd5}
\end{figure}

\subsection{Density Dependence}\label{sec:2rodden}

As the binding density increases, inter-protein interactions have more significant effects on protein binding.
Here, we use the mean-field theory~\cite{tozz21,nogu22,nogu23b}, including orientation-dependent excluded-volume interactions
based on  the theory of Nascimentos et al. for three-dimensional liquid crystals~\cite{nasc17}.
Although 2-rod proteins likely form a smectic liquid crystals at high densities,
we consider only the isotropic and nematic phases in this study.

The free energy $F_{\mathrm {p}}$ of the bound proteins is expressed as follows:
\begin{eqnarray}
F_{\mathrm {p}} &=& \int f_{\mathrm {p}}\ {\mathrm{d}}A, \\ 
f_{\mathrm {p}} &=&  \frac{\phi k_{\mathrm {B}}T}{a_{\mathrm {p}}}\Big[\ln(\phi) + \frac{S \Psi}{2} - \ln\Big(\int_{-\pi}^{\pi} w(\theta_{\mathrm {ps}})\ {\mathrm{d}}\theta_{\mathrm {ps}}\Big)\Big],\hspace{0.5cm} \\
w(\theta_{\mathrm {ps}})  &=&  g\exp\Big[\Psi s_{\mathrm {p}}(\theta_{\mathrm {ps}}) + \bar{\Psi}\sin(\theta_{\mathrm {ps}})\cos(\theta_{\mathrm {ps}}) \nonumber \\
 && - \frac{U_{\mathrm {2rod}}}{k_{\mathrm {B}}T} \Big]\Theta(g), \\
g   &=& 1-\phi (b_0-b_2S s_{\mathrm {p}}(\theta_{\rm ps})),
\end{eqnarray}
where $s_{\mathrm {p}}(\theta_{\mathrm {ps}}) = \cos^2(\theta_{\mathrm {ps}}) - 1/2$ and
$\Theta(x)$ denotes the unit step function.
The order of proteins is obtained by an ensemble average (denoted by angular brackets) of $2s_{\mathrm {p}}$:
\begin{eqnarray} \label{eq:av1}
S &=& 2 \langle s_{\mathrm {p}}(\theta_{\mathrm {ps}}) \rangle,\\
&=&  2\frac{\int_{-\pi}^{\pi} s_{\mathrm {p}}(\theta_{\mathrm {ps}}) w(\theta_{\mathrm {ps}})\ {\mathrm{d}}\theta_{\mathrm {ps}} }{\int_{-\pi}^{\pi}  w(\theta_{\mathrm {ps}}) \ {\mathrm{d}}\theta_{\mathrm {ps}}}, \label{eq:av2}
\end{eqnarray}
where $\theta_{\mathrm {ps}}$ denotes the angle between the major protein axis and ordered direction {\bf S} 
(see Fig.~\ref{fig:cart}).
The factor $g$ expresses the effect of the orientation-dependent excluded volume, where 
 $b_0= (4 + b_{\mathrm {exc}}/2)\lambda$ and $b_2= b_{\mathrm {exc}}\lambda$.
Here, we use $\lambda = 1/3$ and $b_{\mathrm {exc}}= 1.98$ for an elliptic protein with an aspect ratio of $\ell_1/\ell_2=3$,
where $\ell_1$ and $\ell_2$ are the lengths in the major and minor axes, respectively~\cite{nogu22}.
Proteins can have non-overlapping conformations at $g>0$,
and hence, the maximum density is given by a function of $S$ as $\phi_{\rm lim}(S) = 1/(b_0- b_2 S/2)$ (see the rightmost line in Fig.~\ref{fig:dcd0bd0}(b)).
The quantities $\Psi$ and $\bar{\Psi}$ are the symmetric  and asymmetric components of the nematic tensor, respectively, 
and are determined using Eq.~(\ref{eq:av2}) and $\langle \sin(\theta_{\mathrm {ps}})\cos(\theta_{\mathrm {ps}}) \rangle =0$.
Further details of this theory are described in Refs.~\citenum{tozz21} and \citenum{nogu22}.

For the symmetric proteins ($\kappa_{\mathrm{pd}}=C_{\mathrm{pd}}=0$),
the density dependence is qualitatively the same as that for the 1-rod proteins ($\omega=0$) reported in Ref.~\cite{nogu22}.
On a cylindrical membrane with a small curvature of $1/R_{\mathrm {cy}}C_{\mathrm {pm}}\lesssim 0.2$, the 2-rod proteins with $\omega=\pi/6$ exhibit an isotropic--nematic transition at $\phi_{\mathrm {cy}}\simeq 0.11$ (data not shown).
At a middle curvature $0.2 \lesssim 1/R_{\mathrm {cy}}C_{\mathrm {pm}} \leq 1$, the proteins exhibit no phase transition, and the orientational order $S$ increases continuously with increasing $\phi_{\mathrm {cy}}$ (data not shown).
At  $1/R_{\mathrm {cy}}C_{\mathrm {pm}} < 1$, the preferred direction of the proteins is  the azimuthal direction of the membrane tube,
i.e., $\theta_{\mathrm {sc}}=0$.
At  $1/R_{\mathrm {cy}}C_{\mathrm {pm}} \gtrsim 1.3$, the preferred direction is tilted symmetrically to the positive and negative angles, 
as previously explained (see Fig.~\ref{fig:scd0bd0}).
At low densities, proteins with positive and negative preferred angles can coexist at the same amount with keeping $\theta_{\mathrm {sc}}=0$. 
In contrast, at high densities, this coexistence is prevented by the larger excluded-volume interactions between proteins of the different angles.
Second- and first-order phase transitions occur between these two states for middle membrane curvatures ($1/R_{\mathrm {cy}}C_{\mathrm {pm}} < 1.6$) and high membrane curvatures ($1/R_{\mathrm {cy}}C_{\mathrm {pm}} > 1.6$), respectively (see Figs.~\ref{fig:dcd0bd0} and \ref{fig:bd}(a)).
At the first-order transition, the distribution of $\theta_{\mathrm {pc}}$ changes from  two symmetrical peaks to either peak (see the dashed lines in Fig.~\ref{fig:dcd0bd0}(c)), and $\theta_{\mathrm {sc}}$  and $S$ exhibit discrete changes (see Fig.~\ref{fig:dcd0bd0}(a) and (b)).
Conversely, for the second-order transition,  
the two peaks are pushed to $\theta_{\mathrm {pc}}=0$ and unified to reduce the excluded volume before the transition, following which
the single peak continuously moves into either the positive or negative direction above the transition point (see the solid lines in Fig.~\ref{fig:dcd0bd0}(c)). 
In the phase diagram,
the curves of the second- and first-order transitions meet at a single point as shown in Fig.~\ref{fig:bd}(a).
A similar phase diagram is obtained for the 1-rod proteins ($\omega=0$).

For the asymmetric proteins ($\kappa_{\mathrm{pd}} \ne 0$ or $C_{\mathrm{pd}} \ne 0$),
the transition becomes a continuous change; however, a metastable state appears at a high density (see Figs.~\ref{fig:dcd1bd0} and \ref{fig:dcd0bd5}).
At $\kappa_{\mathrm{pd}}=0$ and $C_{\mathrm{pd}}> 0$, the negative angles of $\theta_{\mathrm {pc}}$ have lower bending energies (see Fig.~\ref{fig:scds}(c)),
such that the branch of $\theta_{\mathrm {sc}}<0$ becomes the equilibrium state (see Fig.~\ref{fig:dcd1bd0}).
The other branch becomes the  metastable state that appears at higher membrane curvatures,
and the lower-bound curvature increases with increasing $C_{\mathrm{pd}}$ (see Fig.~\ref{fig:bd}(b)).
Interestingly,
at $\kappa_{\mathrm{pd}}/\kappa_{\mathrm{pm}}=0.5$ and $C_{\mathrm{pd}}=0$,
 the equilibrium value of $\theta_{\mathrm {sc}}$ changes the sign with increasing $\phi_{\mathrm {cy}}$ (see Fig.~\ref{fig:dcd0bd5}(a)).
This is due to high and low peaks at $\theta_{\mathrm {pc}}=\theta_1$ and $-\theta_2$ with $0<\theta_1<\theta_2$ (see the middle solid line in Fig.~\ref{fig:sbds}(c)).
With increasing $\phi_{\mathrm {cy}}$, the lower peak is reduced and subsequently disappears in the equilibrium state (see the solid lines in Fig.~\ref{fig:dcd0bd5}(c)).
Thus, the asymmetry of proteins causes the transition to become a continuous change.
It resembles the aforementioned change from the first-order to continuous change at  $1/R_{\mathrm {cy}}C_{\mathrm {pm}}\simeq 0.2$ in the symmetric proteins.
Note that taking a different protein axis for the elliptical approximation
does not change this binding behavior except for the protein angles.
When the axis of segment $a$ is taken, 
the values of $\theta_{\mathrm {sc}}$ and $\theta_{\mathrm {pc}}$ are shifted by $\omega/2$, while $S$
is unchanged.

\section{proteins of threefold or higher rotational symmetry}\label{sec:three}

Single proteins or protein assemblies often exhibit $N$-fold rotational symmetry with $N\ge 3$.
First, we consider cases with perfect rotational symmetry.
The bending energy of an $N$-fold rotationally symmetric protein is generically expressed as
\begin{eqnarray}
&&U_{\mathrm{r},N}(H, K, D, \theta_{\mathrm {p1}}) = \\ \nonumber
&& \sum_{j=1}^N u_0\Big(H, K, D\cos\big(2(\theta_{\mathrm {p1}}+\frac{2\pi j}{N})\big),
 D\sin\big(2(\theta_{\mathrm {p1}}+\frac{2\pi j}{N})\big)\Big),
\end{eqnarray}
where $K=C_1C_2$ is the Gaussian curvature, 
$u_0(H, K, D\cos(2(\theta_{\mathrm {p1}}+2\pi j/N)), D\sin(2(\theta_{\mathrm {p1}}+2\pi j/N)))$ is the bending energy of 
the $j$-th segment (or protein),
and $\theta_{\mathrm {p1}}$ is the angle between the axis of the first segment and direction of either principal membrane curvature.
Here, we only consider the linear and squared terms,  as is usual for bending energies.
For the symmetry, $U_{\mathrm{r},N}(H, K, D, \theta+2\pi/N)=U_{\mathrm{r},N}(H, K, D, \theta)$.
To satisfy this relation, the linear terms ($\propto \cos(2(\theta_{\mathrm {p1}}+\frac{2\pi j}{N}))$ and $\sin(2(\theta_{\mathrm {p1}}+\frac{2\pi j}{N}))$)
vanish for $N \ge 3$.
The squared terms ($\propto \cos(4(\theta_{\mathrm {p1}}+\frac{2\pi j}{N}))$ and $\sin(4(\theta_{\mathrm {p1}}+\frac{2\pi j}{N}))$)
vanish for $N=3$ and $N \ge 5$, because $e^{ 8\pi{\mathrm {i}}/N}=1$ is satisfied at $N=1$, $2$, and $4$ but otherwise not.
Therefore, for the rotational symmetry of $N=3$ and $N \ge 5$, the bending energy is independent of $\theta_{\mathrm {p1}}$
but is a function of $H$ and $K$, since $D^2= H^2 - K$. 
Hence, it is laterally isotropic, and the Canham--Helfrich energy~\cite{canh70,helf73} is applicable.
For $N=4$, the $\theta_{\mathrm {p1}}$-dependent term remains.
When $u_0= (\kappa_{\mathrm {p}}a_{\mathrm {p}}/2)(H+ D\cos(2(\theta_{\mathrm {p1}}+\frac{2\pi j}{N})) - C_{\mathrm {p}})^2 $ is used,
the protein bending energy is given by
$U_{\mathrm{r},4}(H, K, D, \theta_{\mathrm {p1}}) =\kappa_{\mathrm {p}}a_{\mathrm {p}}[ 2H^2 + D^2(\cos(4\theta_{\mathrm {p1}})+1) + 2C_{\mathrm {p}}^2 ]$.

\begin{figure}[tbh]
\includegraphics[]{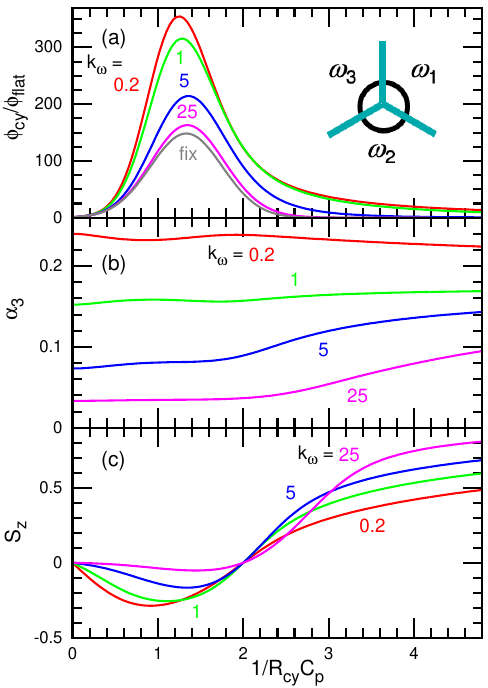}
\caption{
Binding of threefold rotationally symmetric proteins at the low-density limit.
(a) Binding density $\phi_{\mathrm{cy}}$ on a cylindrical membrane  with respect to 
the density $\phi_{\mathrm{flat}}$ on a flat membrane.
Upper four lines: $k_{\omega} = 0.2$, $1$, $5$, and $25$, from top to bottom.
Lowest line: the angles are fixed as $\omega_1=\omega_2=\omega_3=2\pi/3$. 
The schematic of the protein is shown in the inset.
(b) Deformation degree $\alpha_3$ for $k_{\omega} = 0.2$, $1$, $5$, and $25$.
(c) Orientational degree $S_z$ along the ($z$) axis of membrane tube for $k_{\omega} = 0.2$, $1$, $5$, and $25$.
}
\label{fig:3rod}
\end{figure}

Even when a protein has rotational symmetry in its native structure,
the proteins can take asymmetric shapes under protein deformation.
We consider a protein with threefold rotational symmetry, as shown in the inset of Fig.~\ref{fig:3rod}(a).
Three crescent-rod-like segments are connected at the branching point with harmonic angle potentials:

\begin{eqnarray}
U_{\mathrm{3rod}} &=& \sum_{j=1}^3  \frac{\kappa_{\mathrm {p}}a_{\mathrm {p}}}{2}\Big(H+ D\cos\Big(2\big(\theta_{\mathrm {p1}}+\frac{2\pi j}{N}\big)\Big) - C_{\mathrm {p}}\Big)^2  \nonumber \\ 
&&  + \frac{k_{\omega}k_{\mathrm{B}}T}{2}\Big(\omega_j - \frac{2\pi}{3}\Big)^2,
\end{eqnarray}
where $\omega_j$ is the angle between neighboring segments. 
We use $\kappa_{\mathrm {p}}=50k_{\mathrm{B}}T$ and $a_{\mathrm {p}}C_{\mathrm {p}}^2=0.1$.
The protein deformation is quantified by a shape parameter $\alpha_3= \sqrt{\langle (r_{\mathrm{G}}/\ell_{\mathrm{p}})^2\rangle}$,
where $r_{\mathrm{G}}$ is the distance between the center of mass and branching point of the protein,
 and $\ell_{\mathrm{p}}$ is the length of each protein segment.
The orientational order $S_z$ along the ($z$) axis of the membrane tube is given by $S_z= 2(z_{\mathrm{G}}/r_{\mathrm{G}})^2-1$,
where $z_{\mathrm{G}}$ is the $z$ component of the center of mass of the protein (the branching point is the origin of the coordinate).

As the coefficient $k_{\omega}$ of the angle potentials decreases,
the protein exhibits a larger deformation (see Fig.~\ref{fig:3rod}(b))
so that each segment can take its preferred orientation more frequently.
Thus, the binding ratio $\phi_{\mathrm {cy}}/\phi_{\mathrm {flat}}$
increases with decreasing $k_{\omega}$ (see Fig.~\ref{fig:3rod}(a)).
The deformed protein is oriented along the azimuthal and tube axes 
at low and high membrane curvatures, respectively (see Fig.~\ref{fig:3rod}(c)).
Therefore, protein deformation can induce anisotropic bending energy in rotationally symmetric proteins
and enhance curvature sensing.

\section{Summary}\label{sec:sum}

We have studied curvature sensing of proteins with asymmetric shapes and/or protein deformation.
Protein asymmetry breaks the symmetry of sensing with respect to the azimuthal direction on cylindrical membranes,
such that the transition between the symmetrical and asymmetrical angle distributions disappears and the other branch becomes a metastable state.
The $N$-fold rotationally symmetric proteins with $N=3$ or $N\ge 5$ exhibit laterally isotropic bending energies,
when the protein deformation is negligible.
However, their deformation can generate asymmetry in the protein shape and
enhance protein binding to membranes with preferred curvatures.

In this study, we consider the proteins consisting of two rods as asymmetric proteins.
The internal structures affect the  curvature sensing at membrane curvatures higher than their preferred curvatures,
whereas only small modifications occur at lower curvatures.
In general, proteins can have more complicated internal structures.
Hence, the protein bending energy can have nine independent coefficients in Eq.~(\ref{eq:u2}) as
\begin{eqnarray} \label{eq:cvg}
U_{\mathrm {cv}} =&& k_1 H^2 + k_2 H + k_3 K + k_4 D\cos(2\theta_{\mathrm {pc}})  \nonumber \\ \nonumber 
&+& k_5 HD\cos(2\theta_{\mathrm {pc}}) + k_6 D^2\cos(4\theta_{\mathrm {pc}}) \\ \nonumber 
&+& k_7 D\sin(2\theta_{\mathrm {pc}}) + k_8 HD\sin(2\theta_{\mathrm {pc}}) \\
&+& k_9 D^2\sin(4\theta_{\mathrm {pc}}).
\end{eqnarray}
Note that the constant term is neglected, since it can be included in the binding chemical potential as $\mu'= \mu + \mu_{\mathrm {cv}}$.
Isotropic proteins can have the first three terms ($k_1=2\kappa$, $k_2=-2\kappa C_0$, and $k_3 = \bar{\kappa}$ in the Helfrich model~\cite{helf73}).
Twofold rotationally or mirror symmetric proteins can have the first six terms ($k_1$--$k_6$),
and asymmetric proteins can have all terms.
However, it is difficult to determine such many parameters.
The number of parameters should practically be reduced based on each protein structure and experimental/simulation data.

The asymmetry of the protein bending energy can be determined from
the asymmetric angle distribution of bound proteins on symmetrically curved membranes, such as a cylindrical tube.
Currently, it is difficult to measure experimentally.
However, for atomistic and coarse-grained molecular simulations,
binding of a single protein is relatively easy to investigate.
The angle distribution of the protein axis on cylindrical or buckled membranes~\cite{nogu11a,hu13a},
and the curvature sensing of proteins can be evaluated.
A few types of proteins and peptides (amphipathic peptides~\cite{gome16} and F-BAR protein Pacsin1~\cite{mahm19}) have been investigated only on buckled membranes of a single membrane shape.
Protein bending properties
can be more quantitatively evaluated using membranes with various curvatures.
In highly buckled membranes, the membrane curvature under the proteins can vary along the protein axis.
This local curvature difference can also modify curvature sensing.
These protein properties are important for a quantitative understanding of curvature sensing and generation.
Although we focused on the curvature sensing in this study,
the curvature generation should also be modified by the protein asymmetry.

\begin{acknowledgments}
This work was supported by JSPS KAKENHI Grant Number JP21K03481. 
\end{acknowledgments}

\end{document}